# Radiatively broadened thermal emitters

*Simon Huppert,[1] Angela Vasanelli,[1,*] Thibault Laurent,[1] Yanko Todorov,[1] Giulia Pegolotti,[1] Grégoire Beaudoin,[2] Isabelle Sagnes,[2] and Carlo Sirtori[1]*

[1]Université Paris Diderot, Sorbonne Paris Cité, Laboratoire Matériaux et Phénomènes Quantiques, UMR7162, 75013 Paris, France

[2]Laboratoire de Photonique et de Nanostructures, CNRS, 91460 Marcoussis, France

*Corresponding author: angela.vasanelli@univ-paris-diderot.fr

**Abstract**

We study the incandescence of a semiconductor system characterized by a radiatively broadened material excitation. We show that the shape of the emission spectrum and the peak emissivity value are determined by the ratio between radiative and non-radiative relaxation rates of the material mode. Our system is a heavily doped quantum well, exhibiting a collective bright electronic excitation in the mid-infrared. The spontaneous emission rate of this collective mode strongly depends on the emission direction and, uncommonly for a solid-state system, can dominate non-radiative scattering processes. Consequently the incandescence spectrum undergoes strong modifications when the detection angle is varied. Incandescence is modelled solving quantum Langevin equations, including a microscopic description of the collective excitations, decaying into electronic and photonic baths. We demonstrate that the emissivity reaches unity value for a well-defined direction and presents an angular radiative pattern which is very different from that of an oscillating dipole.

Mid-infrared sources are required for many applications such as gas spectroscopy, sensing, imaging, surveillance and threat detection. Efficient semiconductor light-emitting devices are not available in this spectral region due to extremely long spontaneous emission time as compared to non-radiative scattering processes of electrons in solids. Stimulated emission enables to overcome this issue in quantum cascade lasers[1], that are based on a system of tunnel coupled semiconductor quantum wells, engineered to achieve population inversion. However, they do not yet fulfil the needs for broadband and cheap sources. Hence, the most common commercial infrared sources are standard incandescent filaments or membranes.

Quasi-monochromatic infrared thermal sources have been developed making use of Kirchhoff's law, which states that the emissivity of a material equals its absorptivity[2]. The latter can be engineered by structuring the surface of an emitter on a subwavelength scale with periodic gratings[3], photonic crystals[4,5], and metamaterials[6]. This approach requires an accurate electromagnetic modelling of the photonic structure in order to achieve unity absorptivity at the desired wavelength. Quasi-monochromatic thermal emission has also been obtained by exploiting naturally narrow optical resonances of a material, possibly coupled to photonic structures[7,8]. In particular, electronic transitions between confined states in the conduction band of semiconductor quantum wells (QWs) (intersubband transitions), displaying quasi-monochromatic absorption, can be used to realize controlled incandescent devices[5,8,9]. Quantum wells have also been exploited to demonstrate the first dynamic thermal emission control[10].

The design of controlled incandescent devices is based on the maximization of the emissivity through electromagnetic modeling, including non-radiatively broadened material resonances through a semiclassical description[7,8,11]. However, a microscopic description of the material excitations can give a more in-depth insight on the impact of the different decay mechanisms on the incandescent emission. This has been shown for instance in the temperature dependent incandescence of rare earth oxides emitters that can only be explained by taking into account radiative and non-radiative relaxation channels in a microscopic model[12].

In this work we show how the interplay between radiative and non-radiative processes controls the spectral properties of incandescent sources based on highly doped semiconductor quantum wells. Due to the presence of strong carrier-carrier interactions[13,14,15], the absorption spectrum of highly doped quantum wells with several occupied subbands presents a single bright mode, the multisubband plasmon (MSP), concentrating most of the oscillator strength of the system[16,17]. These collective excitations display a superradiant behavior, with a density dependent spontaneous emission rate, that can be evaluated using Fermi's golden rule[18, 19]:

$$\frac{1}{\tau_\theta} = \Gamma_0 \frac{\sin^2 \theta}{\cos \theta} \quad \text{with} \quad \Gamma_0 = \frac{e^2 N_s}{2m^* c \varepsilon_0 \sqrt{\varepsilon_s}} \quad (1)$$

where $N_s$ is the surface electron density, $m^*$ the effective mass, $\varepsilon_s$ the semiconductor dielectric constant and $\theta$ the emission angle measured with respect to the normal to the QW plane. In highly doped GaInAs quantum wells the spontaneous emission time can reach few tens of fs[18]. As this value is much shorter than the characteristic time of any non-radiative scattering event[20], the energy relaxation dynamics can be dominated by the radiative rate $1/\tau_\theta$. In the following we exploit the strong angular dependence of $1/\tau_\theta$ to prove the importance of collective spontaneous emission for plasmon relaxation. Our complete quantum model of the incandescence, based on the resolution of quantum Langevin equations, is then presented. This model, providing a microscopic description of the incandescence without employing Kirchhoff's law or a dielectric function, perfectly reproduces our experimental results. We finally demonstrate that without any photonic structure the emissivity reaches unity value for a well-defined spatial direction as a result of a critical coupling effect, controlled by the matching between spontaneous emission and non-radiative excitation rates.

**RESULTS AND DISCUSSIONS**

Our samples are based on a GaInAs/AlInAs highly doped QW grown by Metal Organic Chemical Vapor Deposition on an InP substrate. The most doped one (named HD) consists of a 100 nm GaInAs layer (embedded between AlInAs barriers) with an electronic density 7.5x10$^{13}$ cm$^{-2}$; the second one is composed of an 18.5 nm QW, with an electronic density 1.2x10$^{13}$ cm$^{-2}$ (named LD). The samples are prepared to perform angle resolved emission measurements, under application of an in-plane current[21]. For this, Ohmic NiAuGe/NiAu contacts are diffused down to the QW and the facet of the sample is mechanically polished with an angle α. A Ti/Au layer is deposited on the top surface of the sample, in order to maximize the electromagnetic field overlap with the QW. Figure 1a presents a sketch of the sample geometry with a schematic representation of MSP propagation in the QW plane, characterized by an in-plane wavevector $\vec{k}$. MSPs are thermally excited by the application of an in-plane current modulated with a square wave at 10 kHz and decay by emitting photons with the same in-plane wavevector $\vec{k}$. All measurements are performed at room temperature.

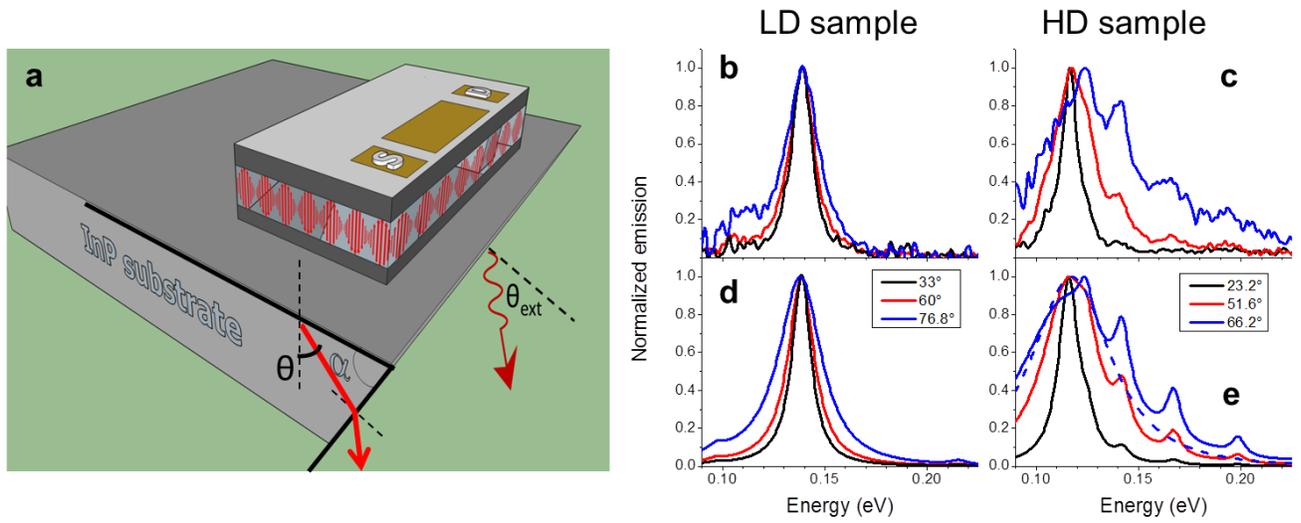

Figure 1: **(a)** Sketch of the device for thermal emission of multisubband plasmons. An in-plane current is injected between the source (S) and the drain (D) of the device. Thermal radiation is extracted from a polished facet (α is the polishing angle). The internal angle for light propagation is indicated with θ, while $θ_{ext}$ represents the external angle. On the side of the mesa we have schematically represented the propagation of the plasmon in the layer plane. Panels **(b)** and **(c)** present the normalized measured emission spectra of the two samples at different θ, compared with the simulated ones (shown in panels **(d)** and **(e)**, solid lines). Dashed line in panel (e) corresponds to single plasmon approximation of the spectrum at 66.2°.

Figure 1 presents simulated and experimental plasmonic incandescence spectra for different emission angles and doping densities. Figures 1b and 1c show normalized spectra obtained respectively for samples LD and HD at three different values of the internal emission angle θ. At low θ, the spectra are quasi-monochromatic, with a Lorentzian shape, centered at the energy of the MSP, $\hbar\omega_{MSP}$. The full width at half maximum is 10 meV for LD (respectively 8 meV for HD), and it is dominated by non-radiative phenomena. When increasing θ, the linewidth gradually increases in a manner that is well described by equation (1). This radiative broadening indicates that photon emission is the most efficient relaxation mechanism for MSPs. Figure 1c also shows resonances at higher energy, appearing for large values of θ; they are due to secondary plasmon modes which will be discussed further.

It is important to underline that the radiative lifetime varies from 1.4 ps at 20° to 69 fs at 70° for sample LD (respectively from 250 fs to 12 fs for HD). These dramatic variations not only affect the linewidth, but also control the emissivity and the radiation directionality in a way that can only be interpreted by accounting consistently for both radiative and non-radiative decay of MSPs. In order to characterize accurately the angular dependence of MSP incandescence, we have measured angle-resolved radiated spectra. Three different polishing angles have been used, in order to cover values of θ ranging from 25° to 85°. Panels (a) and (b) of Fig. 2 present the emitted power (in color scale) as a function of the photon energy and emission angle. The emission map has a droplet shape with maximum emission at an angle of ~55° in LD sample and ~35° in HD sample. A similar conclusion can be drawn by plotting the total emitted power integrated over the detection bandwidth, Fig. 2e and 2f (square symbols): the two samples have distinct preferential emission directions.

This behavior is in apparent contradiction with Eq. (1) which naturally would provide an evaluation of the emitted power per unit solid angle as:

$$dP = \frac{\hbar\omega_{MSP}}{\tau_\theta} n_B\left(\hbar\omega_{MSP}, T\right) dN_{pl} \qquad (2)$$

with $n_B(\hbar\omega_{MSP}, T)$ the Bose-Einstein distribution for plasmon occupancy at energy $\hbar\omega_{MSP}$ and temperature T, while dN$_{pl}$, proportional to cosθ dΩ, is the number of plasmon modes such that their emission direction at resonance is contained in the solid angle of detection dΩ. Equations (1) and (2) would yield a dipole-like emission described by Larmor's formula, i.e. sin²θ dependence with maximum at θ = 90°, which does not agree with our observations. Indeed equation (2) is obtained assuming that all plasmons thermalize at temperature T. However, thermal equilibrium fails to establish when the collective mode is mostly radiatively broadened (hence radiatively cooled down). We will show in the following that the angular dependence of the emission results from the interplay between radiative and non-radiative damping, which is captured by computing the incandescence spectra through a twofold procedure: in a first step the electron-electron interaction is numerically diagonalized to determine the bright MSP modes[17]. These include the main MSP discussed above and secondary modes with much lower oscillator strength. In a second step we solve quantum Langevin equations in an input-output formalism to calculate the emission[22-25] (more details on the model can be found in the supplementary material). As it can be seen in Figs. 1d and 1e, this method reproduces accurately all the experimental features observed, in particular the angle-dependent broadening, and the appearance of extra peaks beyond θ = 40° in sample HD. These additional resonances are associated with high energy plasmon modes issued from the dipole – dipole coupling among intersubband transitions between non-consecutive levels[17]. It is remarkable that these modes are clearly visible (see also Figs 1c and 1e) although their oscillator strength is more than ten times lower than that of the main MSP. The reason for this is that the main emission peak tends to vanish at high θ, while the secondary modes gain in brightness. Figs. 2c and 2d present calculated emission color maps which accurately reproduce the directional emission of both samples (the emission maxima are indicated by arrows in Fig. 2).

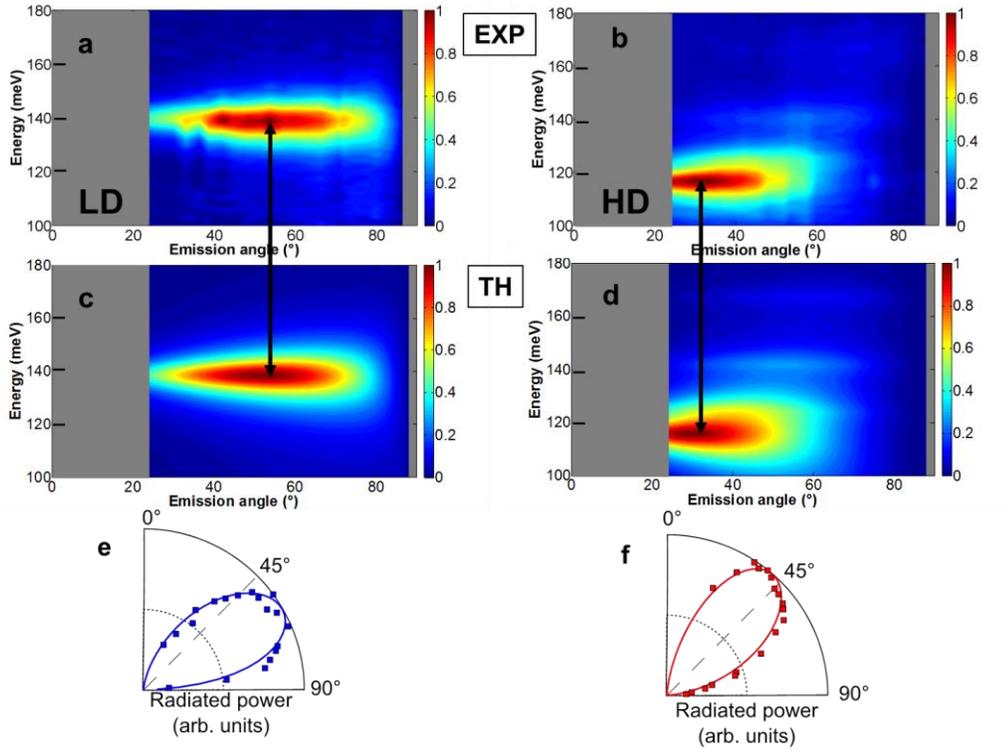

**Figure 2:** Measured (panels (a) and (b)) and simulated (panels (c) and (d)) color maps of the emitted radiation (in arbitrary units) as a function of the photon energy and of the internal angle θ, for the two samples LD and HD. The black arrows indicate the preferential angle of emission. Panels (e) and (f) present a polar plot of the integrated emitted power as extracted from the measured (squares) and simulated (lines) angle resolved spectra, respectively for samples LD and HD.

Analytical results obtained in single plasmon approximation (discarding all secondary MSP modes) provide in-depth insights on the physical origin of the directionality of MSP incandescence. In our formalism thermal emission arises from the coupling of the main MSP with two reservoirs: a hot bath of electronic oscillators, responsible for both thermal excitation of plasmons and their non-radiative decay, and a cold bath of photons. The MSP couples to the electronic reservoir through a constant damping rate γ, that can be determined from low θ measurements to be 8 and 10 meV for HD and LD respectively. Conversely the coupling to the photon bath is characterized by a frequency and angle dependent rate[22]

$$\Gamma_\theta(\omega) = \frac{\omega}{\omega_{MSP}\tau_\theta}$$ . We consider an input of thermal excitations in the hot bath at temperature T. The

emitted intensity is then calculated from the output photon occupancy, expressed by the following formula:

$$\left\langle a_{q,\vec{k}}^{out\dagger} a_{q,\vec{k}}^{out} \right\rangle = \alpha_\theta(\omega) n_B(\hbar\omega, T)$$

$$\alpha_\theta(\omega) = \frac{\dfrac{4\omega_{MSP}^2}{(\omega_{MSP}+\omega)^2} \gamma \, \Gamma_\theta(\omega)}{(\omega_{MSP}-\omega)^2 + \dfrac{4\omega_{MSP}^2}{(\omega_{MSP}+\omega)^2}\left(\dfrac{\gamma}{2}+\dfrac{\Gamma_\theta(\omega)}{2}\right)^2} \quad (3)$$

In this relation, $\omega$ denotes the frequency of the output photon mode, described by the operator $a_{q,\vec{k}}^{out}$ (with $q$ the component of the photon wavevector along the growth direction). Anti-resonant interaction terms are included in the calculation and are responsible for the factor $\dfrac{4\omega_{MSP}^2}{(\omega_{MSP}+\omega)^2}$ in Eq. (3). This term is close to 1 near the resonance, but it becomes significant out of resonance when the radiative broadening is large. Note that $\alpha_\theta(\omega)$ corresponds to the absorbance of the MSP that can be analytically calculated by considering a photonic input in our model. This is exactly what is expected from Kirchhoff's law of thermal emission, namely that the incandescent flux is the product of the absorbance times the spectral radiance of a perfect black-body at the same temperature[2].

For fixed values of the dimensionless parameters $\gamma/\omega_{MSP}$ and $kT/\omega_{MSP}$, the incandescence spectrum described by equation (3) is a universal function of the coupling ratio $g_\theta = \Gamma_\theta(\omega_{MSP})/\gamma$, that can be decomposed as $g_\theta = g_0 \, f(\theta)$ where $g_0$ is proportional to the electronic density $N_s$ and $f(\theta) = \dfrac{\sin^2\theta}{\cos\theta}$.

Figure 3 presents the emission spectra calculated for different values of $g_\theta$ as a function of the normalized frequency $\omega/\omega_{MSP}$ (the temperature is such that $kT/\omega_{MSP}=0.35$). Plasmon incandescence is also compared to the blackbody spectrum (black line). Panels (a), (b) and (c) correspond to three different regimes, that can be spanned by changing the detection angle.

For small values of $g_\theta$ (Fig. 3a), i.e. for low $g_0$ or $\theta$ close to 0°, the spectrum is Lorentzian with a width $\gamma/\omega_{MSP}$ and an amplitude proportional to $g_\theta$. In this case the power radiated in a solid angle is given by eq. (2). The plasmon population is at equilibrium with the electronic bath and only the spontaneous emission rate determines the emitted power: the bottleneck phenomenon is in this case the coupling to the photonic bath ("photon bottleneck" regime). The radiated power at a given θ in this regime is proportional to the electronic density in the QW, in agreement with eq. (1).

The maximum of $\alpha_\theta(\omega)$ is achieved for $g_\theta =1$ (Fig. 3b), where the MSP emissivity reaches 1 and the emission spectrum coincides at the frequency $\omega_{MSP}$ with that of a perfect blackbody. This situation corresponds to a "critical coupling" regime of incandescence: MSPs are excited at the same rate as they decay radiatively. The plasmon thus plays the role of a spectral filter for the blackbody emission, funnelling the electronic excitations of the bath towards free space photons.

For $g_\theta >1$ (Fig. 3c), i.e. when spontaneous emission is the dominant relaxation mechanism for the MSP, the linewidth is significantly increased while the peak power decreases. The spectra are deformed from their Lorentzian shape due to several combined effects: the frequency dependence of the Bose-Einstein distribution and of the absorbance and the inclusion of the anti-resonant interaction terms, which increase $\alpha_\theta(\omega)$ below $\omega_{MSP}$ and reduce it above the resonance. The role of the anti-resonant terms, similar to that observed in ultra-strongly coupled systems[22,26] will be examined in more detail in further work. Note that the area under the spectra presented in Fig. 3c, obtained for different values of $g_\theta$, does not vary linearly with $g_\theta$, as it was the case in the opposite limit $g_\theta<1$. Indeed, the spontaneous emission time is so short that the MSP population is limited by the thermal excitation rate γ and in this case the bottleneck mechanism is the coupling to the electronic bath ("electronic bottleneck" regime). The radiated power in solid angle dΩ is then approximately expressed as $dP_{g_\theta>1} = \gamma\hbar\omega_{MSP}n_B(\hbar\omega,T)dN_{pl}$, which tends to zero for θ→ 90° due to the cosθ dependence of $dN_{pl}$. This explains why the emission is never maximal at 90°, despite the divergence of $g_\theta$.

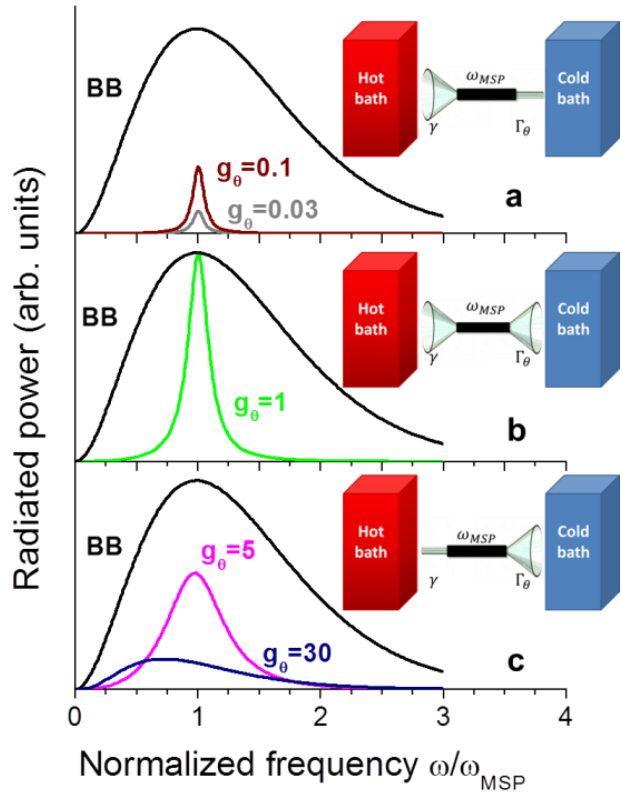

**Fig. 3**: Emission spectra calculated for different values of $g_\theta$ as a function of the normalized frequency $\omega/\omega_{MSP}$ at a temperature such that $kT/\omega_{MSP} = 0.35$ and $\gamma/\omega_{MSP} = 0.05$. Plasmon incandescence is compared to the blackbody spectrum (black line). Panels (a), (b) and (c) correspond to the three different regimes described in the main text ((a) photon bottleneck, (b) critical coupling and (c) electronic bottleneck).

Figure 4a shows the angular dependence of $g_\theta = g_0\, f(\theta)$ for different values of $g_0$ corresponding to different electronic densities. Due to the divergence of $f(\theta)$, the critical coupling condition, with unitary emissivity, is always met at a certain angle $\theta_c$ which decreases when $g_0$ increases. Beyond $\theta_c$, the peak emissivity falls down, as illustrated in Fig. 4b for different values of $g_0$. Finally Fig. 4c presents the angular dependence of the integrated radiated power for different values of $g_0$. For $g_0 \ll 1$ (low electronic density), the system stays in the photon bottleneck regime until the emission angle is close to 90°. As expected, the radiation angular profile tends to that of a dipole, described by Larmor's formula, i.e. proportional to $\sin^2\theta$ (dashed line). For intermediate densities, the preferential emission direction varies with $g_0$, getting closer to normal emission for $g_0 \gg 1$, in which case the system stays in the electronic bottleneck regime for almost all values of $\theta$. Therefore the directionality of the plasmonic incandescence is intrinsically related to the existence of the electronic bottleneck regime induced by the very short spontaneous emission time of the superradiant MSP excitations.

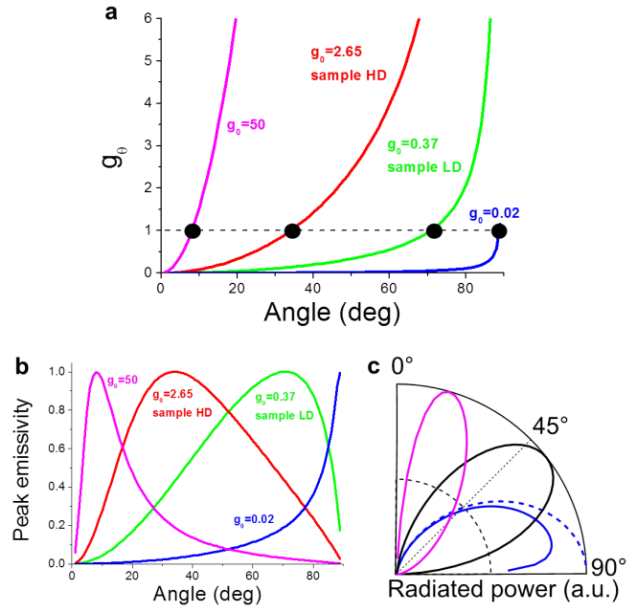

**Fig. 4:** (a) Calculated angular dependence of $g_\theta$ for different values of $g_0$ (i.e. for different electronic densities). The black bullets indicate the values for which the critical coupling condition is met. (b) Angular dependence of the peak emissivity calculated for the same values of $g_0$ as in panel (a). The value 1 corresponds to critical coupling. (c) Angular dependence of the integrated radiated power for different values of $g_0$ (solid lines): $g_0$=50 (pink), $g_0$=1 (black), $g_0$=0.02 (blue). The dashed blue line represents the radiated emission of an oscillating dipole, following Larmor's formula.

The results presented in this work open entirely new possibilities for designing infrared sources of radiation. The combination of the concepts presented here with the quantum engineering of collective excitations[17] allows spectral and directional control of incandescence. In addition, owing to the very fast temperature response of the electron gas, such sources can be modulated at very high rates[10]. Furthermore, within our theoretical framework the statistical properties of the electronic excitations are naturally related to the quantum fluctuations of the emitted light[27], which enables quantum-optical investigation of infrared sources. Finally, beyond the context of thermal emission, our model also allows treating the cases of narrowband optical or electronic inputs[22].

**Acknowledgments.**

We acknowledge financial support from ERC (grant ADEQUATE), Labex SEAM, Renatech Network and Agence Nationale de la Recherche (grant ANR-14-CE26-0023-01).We thank Jean-Jacques Greffet for fruitful discussions.

**Supporting Information Available**.
(1) Details on the theoretical model used to compute the incandescence spectra and derivation of equation 3.
(2) Details on the method used to isolate the plasmonic contribution to the measured incandescence.